\documentclass[a4paper,twocolumn,11pt,accepted=2025-03-20]{quantumarticle}
\pdfoutput=1
\usepackage[utf8]{inputenc}
\usepackage[english]{babel}
\usepackage[T1]{fontenc}
\usepackage{amsmath}
\usepackage{hyperref}
\usepackage{braket}

\usepackage{listings}

\usepackage{amsmath}
\usepackage{amssymb}
\usepackage{hyperref}
\usepackage[resetlabels]{multibib}

\usepackage{tikz}
\usepackage{lipsum}

\newcites{article}{article references}
\newcites{book}{book references}
\newcites{misc}{misc references}
\newcites{repo}{repository references}
\newcites{web}{website references}
\newcites{other}{Other reference}

\begin{document}

\title{Ring-exchange physics in a chain of three-level ions}

\author{Sourav Biswas}
\affiliation{DIPC - Donostia International Physics Center, Paseo Manuel de Lardiz{\'a}bal 4, 20018 San Sebasti{\'a}n, Spain}
\email{sourav.biswas@dipc.org}
\author{E. Rico}
\affiliation{DIPC - Donostia International Physics Center, Paseo Manuel de Lardiz{\'a}bal 4, 20018 San Sebasti{\'a}n, Spain}
\affiliation{EHU Quantum Center and Department of Physical Chemistry, University of the Basque Country UPV/EHU, P.O. Box 644, 48080 Bilbao, Spain}
\affiliation{European Organization for Nuclear Research (CERN), Geneva 1211, Switzerland}
\affiliation{IKERBASQUE, Basque Foundation for Science, Plaza Euskadi 5, 48009 Bilbao, Spain}
\email{enrique.rico.ortega@gmail.com}
\author{Tobias Grass}
\affiliation{DIPC - Donostia International Physics Center, Paseo Manuel de Lardiz{\'a}bal 4, 20018 San Sebasti{\'a}n, Spain}
\affiliation{IKERBASQUE, Basque Foundation for Science, Plaza Euskadi 5, 48009 Bilbao, Spain}
\email{tobias.grass@dipc.org}
\maketitle

\begin{abstract}
In the presence of ring exchange interactions, bosons in a ladder-like lattice may form the bosonic analogon of a correlated metal, known as the d-wave Bose liquid (DBL). In this paper, we show that a chain of trapped ions with three internal levels can mimic a ladder-like system constrained to a maximum occupation of one boson per rung. The setup enables tunable ring exchange interactions, transitioning between a polarized regime with all bosons confined to one leg and the DBL regime. The latter state is characterized by a splitting of the peak in the momentum distribution and an oscillating pair correlation function.
\end{abstract}

\maketitle   
\section{Introduction}
Metallic behavior is common in electronic matter: Fermionic particles partially fill an energy band. However, if matter is bosonic instead of fermionic, the particles are either expected to condense into a state with superfluid properties or to form an insulating phase~\cite{Fisher_1989}. The possibility of an intermediate phase in which bosonic quasiparticles exhibit metal-like behavior has long been a vivid research subject~\cite{Phillips_2003, Yang_2019, Hegg_2021}. Bosonic lattice models that support metallic behavior have been proposed in Refs.~\cite{Paramekanti_2002,Motrunich_2007}, with ring exchange interactions on a plaquette being the key ingredient, that is, two-body processes, in which two particles residing on a lattice plaquette simultaneously hop along opposite directions.
A quasi-one-dimensional variant of such a model consists only of two chains coupled via a ring exchange term, and potentially interchain hopping~\cite{Sheng_2008}, or extensions to multi-leg ladders \cite{Mishmash_2011,Block_2011}. As a striking consequence of the ring exchange term the occurrence of an unusual strong-coupling phase of bosons has been reported, characterized by a peak splitting in the momentum distribution and sign oscillations of pair correlations, hinting towards the $d$-wave correlated nature of this phase which has been dubbed $d$-wave Bose liquid (DBL). This is an example of emergent metallic behavior in a system of bosons, which cannot be perceived from the physics of superfluidity. It extends the understanding of interacting bosons beyond the boundaries of characterization of phases of matter in terms of spontaneous symmetry breaking of local order parameters.

Realizing prototypical models in controllable quantum systems has become possible through the development of various quantum simulation platforms. Cold atoms in optical lattices have a long-standing history of being used to study many-body phases of bosons~\cite{Greiner_2002}, and this platform has also allowed for realizing systems in ladder geometries~\cite{Atala_2014}. However, the implementation of models with ring exchange interactions remains outstanding. 

Bosonic lattice models have also been realized using excitons in artificial lattices \cite{Lagoin_2022BH,Lagoin_2022}, and the dipolar nature of excitons might even provide weak ring exchange terms, but so far this platform has mostly been used to study gapped phases. 
Another quantum simulation platform that is very powerful for the study of one-dimensional spin models is trapped ions~\cite{Monroe_2021}. An XX chain of spin-1/2 degrees of freedom, $H_{XX}=  \sum_{i,j} J_{ij} \sigma_i^+ \sigma_{j}^- +{\rm h.c.}$, with $\sigma^{\pm}_i$ being raising/lowering operators of spin $i$, has been realized with trapped ions in Ref.~\cite{Jurcevic_2014}, and is equivalent to a model of hard-core bosons with long-range hopping $-J_{ij}$ along a chain. It has also been proposed to use the presence of nearest- and next-nearest neighbor hopping to map the chain onto a two-leg ladder~\cite{Grass_2015}. Another interesting aspect of ion chains is the possibility of going beyond spin-1/2 physics by exploiting three or more internal levels. This can give rise to SU(3) physics, as proposed in Ref.~\cite{Grass_2013}, or spin-1 models as realized in Ref.~\cite{Senko_2015}, or qudit quantum computers as realized in Ref.~\cite{Ringbauer_2022}.

In the present paper, we exploit the possibility of using three-level trapped ion systems for simulating a bosonic two-leg ladder constrained to a maximum occupation of one particle per rung. Appropriately chosen Raman couplings between the levels provide, within a second-order Magnus expansion, the analog of tunable intra-leg tunneling and ring exchange terms. We show that the tunable strength of the ring exchange term leads to a quantum phase transition into the DBL phase and that measuring pairwise correlations between the ions provides clear signatures of this transition. For weak ring exchange terms, the long-range nature of the ion systems becomes important, polarizing the system into one ladder. Our theoretical study employs the (quasi-)exact numerical methods of diagonalization (ED) and density matrix renormalization group (DMRG) through the ITensor library \cite{itensor}. The DMRG algorithm \cite{DMRG_White,Scholl_MPS} has been an extremely reliable and effective tool for investigating different phases of matter in ladder-like systems \cite{Mishra_2012,Mishra_2013,Mishra_2014,Dalmonte_Lad_2023,Gia_Lad_2023,Luca_Lad_2023}. 

\begin{figure*}[t]
    \centering
    \includegraphics[width=0.98\textwidth]{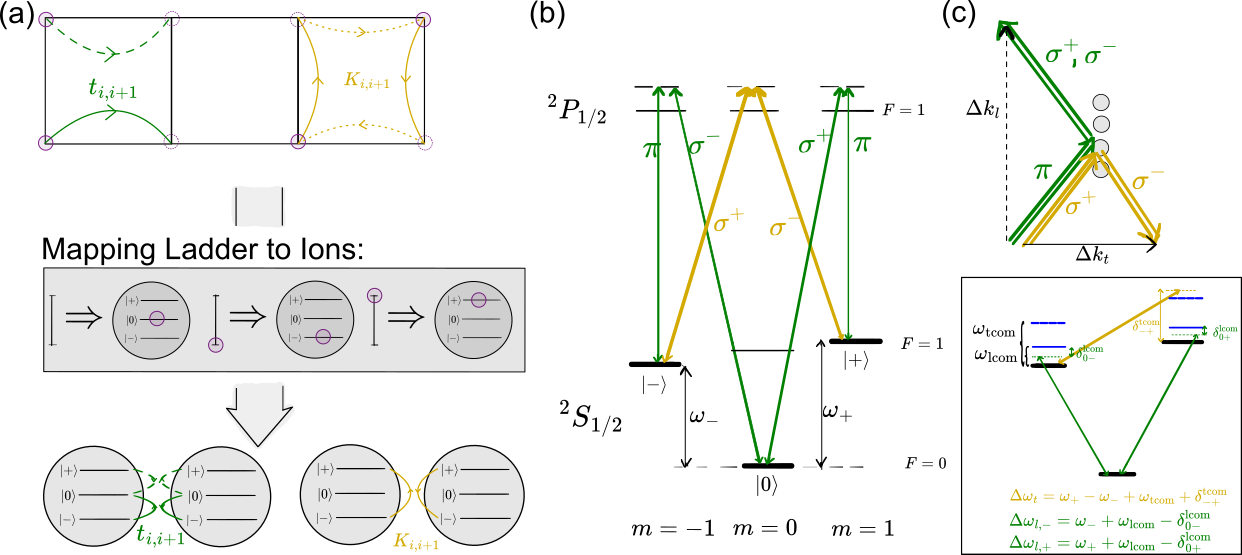}
    \caption{(a) Mapping between the two-leg ladder and chain of three-level ions: Each rung of the ladder is represented by an ion. A hopping process $t_{i,i+1}$ on the lower leg of the ladder (green solid line) maps onto the combined transition of two ions, from $\ket{-}$ to $\ket{0}$ on ion $i$, and from $\ket{0}$ to $\ket{-}$ on ion $i+1$. Similarly, hopping on the upper leg (green dashed line) translates into transitions between $\ket{+}$ to $\ket{0}$. The ring exchange interaction describes correlated hopping processes of two particles on opposite corners of a plaquette (yellow lines). These processes translate into transitions between $\ket{-}$ to $\ket{+}$ on two neighboring ions.
    (b,c) Implementation of hopping and ring exchange via Raman transitions following the level diagram of $^{171}{\rm Yb}^+$ shown in (b). As illustrated in (c), the pairs drawn in green couple to the longitudinal phonons, and are equipped with two beat notes, tuned to the $\tau_{0-}$ transition and the $\tau_{0+}$ transition, with different detunings $\delta_{0-}^{\mathsf{lcom}}$ and $\delta_{0+}^{\mathsf{lcom}}$ from the longitudinal center-of-mass mode at frequency $\omega_{\mathsf{lcom}}$, as indicated in the box. The pairs drawn in yellow couple to transverse phonon modes, and are equipped with one beat-note tuned to the $\tau_{-+}$ transition with a detuning $\delta_{-+}^{\mathsf{tcom}}$ from the transverse center-of-mass mode at frequency $\omega_{\mathsf{tcom}}$. Both couplings operate on the blue sideband.
    \label{fig:scheme}}
    \label{fig:enter-label}
\end{figure*}

\section{Model} 
Let us start with the ring-exchange model on a two-leg ladder studied in Ref.~\cite{Sheng_2008}. Denoting by $a_{i\sigma}^\dagger (a_{i\sigma})$ the creation (annihilation) operators of hard-core bosons on the sites of a ladder of length $L$, identified by a rung index $i \in [1, L]$ and a leg index $\sigma \in \{\downarrow,\uparrow\}$, the model reads
\begin{align}
H = & \sum_{j>i} \left( \sum_\sigma  -t_{ij}^\sigma a_{j\sigma}^\dagger a_{i\sigma}  + K_{ij} a_{i\downarrow}^\dagger a_{j\uparrow}^\dagger a_{j\downarrow} a_{i\uparrow}  \right) + {\rm h.c.}.
\label{Eq:Hamgen}
\end{align}
 In this formulation, we have accounted for a possible long-range character of intra-leg hopping $t_{ij}^\sigma$ in leg $\sigma$, and ring-exchange interactions $K_{ij}$. In the original model of Ref.~\cite{Sheng_2008}, these terms reduce to nearest neighbor (NN) terms, $t_{ij}^\sigma = t \delta_{j,i\pm1}$ and $K_{ij} = K \delta_{j,i\pm 1}$, and the hopping amplitudes are equal in both legs. We also note that $H$ in Eq.~(\ref{Eq:Hamgen}) does not contain an inter-leg hopping term, $H_{\uparrow\downarrow}=-t_{\uparrow\downarrow} \sum_i a_{i\uparrow}^\dagger a_{i\downarrow} + {\rm h.c.}$, and hence the population of each leg is conserved. Although inter-leg hopping could be added to the setup proposed below, we are not interested in this term, as it diminishes the extent of the Bose metal phase, see Ref.~\cite{Sheng_2008}.

To map the Hamiltonian Eq.~(\ref{Eq:Hamgen}) onto a chain of three-level ions, we extend the hardcore constraint from on-site to on-rung, that is, we allow for a maximum occupation of one boson per rung, $a_{i\sigma}^2 = a_{i\uparrow} a_{i\downarrow}=0$. With this constraint, the local Hilbert space on each rung is limited to three states, which we denote by $|-\rangle$ for the lower leg ($\sigma=\downarrow$) being occupied, $|+\rangle$ for the upper leg ($\sigma=\uparrow$) being occupied, and $|0\rangle$ for an empty rung. As illustrated in Fig.~\ref{fig:scheme}(a), the ladder can then be mapped onto a chain of three-level ions. The chain of ions can be interpreted as an array of interacting spin-one objects. However, various non-bilinear terms appear if the Hamiltonian is written in terms of three spin-one operators, spanning the SU(2) algebra. Instead, if we formulate the Hamiltonian in terms of the eight generators of the SU(3) isospin algebra SU(3)  \cite{Gell-Mann1962}, then the Hamiltonian is quadratic and offers convenience of implementation and analysis. As an overcomplete basis of these generators, let us introduce the SU(3) operators $\tau^i_{\alpha\beta} \equiv |i,\alpha\rangle \langle i,\beta |$, with $\alpha,\beta \in \{-,0,+\}$ denoting the internal level of ion $i \in [1, L]$. A hopping process in the lower leg, $a_{i+1\downarrow}^\dagger a_{i\downarrow}$, is re-written as $\tau^{i+1}_{-0}\tau^i_{0-}$, and similarly, $a_{i+1\uparrow}^\dagger a_{i\uparrow} = \tau^{i+1}_{+0}\tau^i_{0+}$ for hopping in the upper leg. The ring-exchange term takes an equally simple quadratic form, $a_{i \uparrow}^\dagger a_{i+1 \downarrow}^\dagger a_{i+1\uparrow} a_{i\downarrow} = \tau^i_{+-}\tau^{i+1}_{-+}$.

\section{Implementation}
The SU(3) model can be engineered with trapped ions, and our proposal closely follows the implementation of spin-1 physics realized in Ref.~\cite{Senko_2015}. We propose to exploit three atomic levels
$\ket{-} = \ket{F=1,F_z=-1}$, $\ket{0} = \ket{F=0,F_z=0}$,  and $\ket{+} = \ket{F=1,F_z=+1}$, all within the $^{2}S_{1/2}$ shell of $^{171}{\rm Yb}^+$. 
The bare ion Hamiltonian in the lab frame is $H_0= \hbar\omega_- |-\rangle \langle -| + \hbar\omega_+ |+\rangle \langle +|$, with $(\omega_++\omega_-)/2$ the hyperfine frequency between the $F$ shells, and $(\omega_+-\omega_-)$ the Zeeman splitting between $F_z=\pm1$.
The desired SU(3) couplings are implemented via  Raman couplings to the $^{2}P_{1/2}$ shell, as illustrated in the level scheme of Fig.~\ref{fig:scheme}(b). According to our mapping between the ladder system and three-level ions, the coupling between $\ket{-}$ ($\ket{+}$) and $\ket{0}$ corresponds to a creation/annihilation process on the lower (upper) leg \footnote{For brevity, we drop the ion index $i$ when it is not needed.}. Detuning of this coupling avoids first-order processes, but second-order processes, combining raising and lowering transitions on two ions, become resonant. Therefore, in leading order, the Raman coupling engineers the equivalent of hopping processes within one leg of the ladder. Similarly, the coupling between $\ket{-}$ and $\ket{+}$, which in first order would correspond to a hopping process along the rung, $a_\downarrow^\dagger a_\uparrow$ or $a_\uparrow^\dagger a_\downarrow$, leads to ring-exchange interactions in a second order. 

A formal derivation of the effective second-order Hamiltonian is provided in the appendix. 
Its general form is $H_{\rm eff}=H_{\rm eff}^{0-}+H_{\rm eff}^{0+}+H_{\rm eff}^{-+}$, where $H_{\rm eff}^{\alpha\beta}$ denotes the effective Hamiltonian stemming from an individual Raman coupling between $\ket{\alpha}$ and $\ket{\beta}$:
\begin{align}
\label{Heff}
H_{\rm eff}^{\alpha\beta} = & \sum_{i\neq j}  J_{\alpha\beta}^{ij} \tau_{\alpha\beta}^i \tau_{\beta\alpha}^j + \sum_{i} V_{\alpha\beta}^i (\tau_{\alpha\alpha}^i-\tau_{\beta\beta}^i).
\end{align}
Importantly, we note that interference between different Raman couplings is avoided by choosing different detunings from the phonon sideband.
The absence of interference becomes evident from the derivation of Eq.~(\ref{Heff}), see appendix.

The amplitudes in Eq.~(\ref{Heff}) are given by:
\begin{align}
J_{\alpha\beta}^{ij} =  \frac{1}{8}\sum_{\mathsf{m}} \eta_{\mathsf{m},i}\eta_{\mathsf{m},j}\frac{\hbar \Omega_{\alpha\beta}^2}{\delta_{\alpha\beta}^{\mathsf{m}}},
\label{Jij}
\end{align}
and
\begin{align}
    V_{\alpha\beta}^i = \frac{1}{4} \sum_{\mathsf{m}} \frac{\Omega_{\alpha\beta}^2}{ \delta_{\alpha\beta}^{\mathsf{m}}} \eta_{\mathsf{m},i}^2 n_{\mathsf{m}}.
    \label{Vi}
\end{align}
Here, $\Omega_{\alpha\beta}$ is the Rabi frequency of the coupling, $\eta_{\mathsf{m},i}$ is the Lamb-Dicke parameter, specifying the displacement of ion $i$ in phonon mode $\mathsf{m}$, and $\delta_{\alpha\beta}^{\mathsf{m}}$ is the detuning of the coupling from the side band of mode $\mathsf{m}$. For $\{\alpha,\beta\}=\{0,+\}$ and $\{\alpha,\beta\}=\{0,-\}$, the first term in Eq.~(\ref{Heff}) describes hopping processes in the upper and lower leg from $i$ to $j$, $t_{ij}^\uparrow = -J_{0+}^{ij}$ and $t_{ij}^\downarrow = -J_{0-}^{ij}$. From $\{\alpha,\beta\}=\{-,+\}$, we obtain the exchange interactions, $K_{ij} = J_{-+}^{ij}$.

\begin{figure}[t!]
    \centering
    \includegraphics[width=0.45\textwidth]{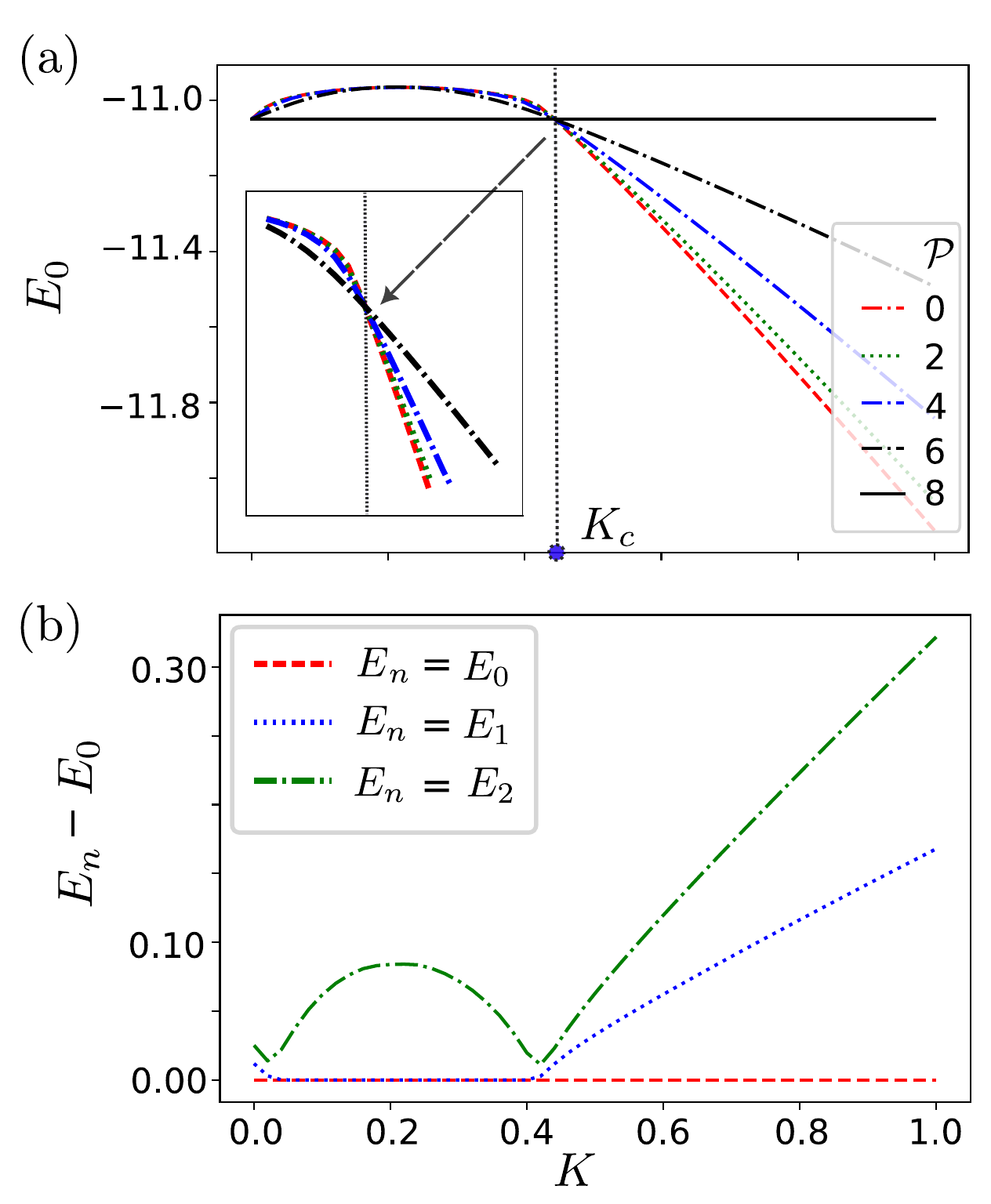}
    \caption{Eigen-energies, computed using ED (for $L=16$), are shown as a function of $K$. Panel (a) depicts the lowest energies for various polarization sectors. In the inset, we zoom into the transition point, noting that the order of the different polarization sectors is altered. Panel (b) shows the energy difference between different lowest-lying energies in $\mathcal{P}$=0. For intermediate values of $K$, a degeneracy is observed, which is lifted when the system enters the DBL phase at large $K$. The degeneracy is also lifted at $K \to 0$, where the system becomes superfluid \cite{Sheng_2008}.} \label{fig:Ens}
\end{figure}

The second term in Eq.~(\ref{Heff}), $\sim V_{\alpha\beta}^i$, can be interpreted as local chemical potential. We can neglect this term for the following reasons: (i) the potential difference between legs is irrelevant since occupation number of each leg is conserved due to the absence of inter-leg tunneling ($t_{\uparrow\downarrow}=0$); (ii) the dependence of the potentials on $i$ is weak since the main contribution is Eq.~(\ref{Vi}) stems from the least detuned center-of-mass mode, with $\eta_{\mathsf{com},i}={\rm const.}$;
(iii) the overall strength of $V_{\alpha\beta}^i$ is proportional to the occupation number $n_{\mathsf{m}}$ of the modes, which is zero in the motional ground state. If the system is initially cooled into the motional ground state, the steady state of a far-detuned Raman coupling remains close to the motional ground state.

It is important to note that all $J_{\alpha\beta}^{ij}$ have a long-range character whose range can be controlled by the detuning of the coupling: If the detuning is close to resonance with the center-of-mass phonon, $J_{\alpha\beta}^{ij}$ would barely decay. Detuning further away from the phonons leads to an (approximately) algebraic decay, $J_{\alpha\beta}^{ij}\sim|i-j|^{-\nu}$. Here, $\nu=3$ is a theoretical upper bound in the limit of an infinitely far detuning from the phonon spectrum. In the following, we will only assume that the decay is sufficiently fast to suppress long-range terms, such that we can limit ourselves to the dominant NN terms and sub-leading next-nearest neighbor (NNN) terms. With this assumption, we can capture the main physics of the system, while the DMRG method, best suited for local Hamiltonians, is still applicable. An ED study of the untruncated model is provided in the appendix (Sec. B).
In terms of the original model, Eq.~(\ref{Eq:Hamgen}), this truncation means that from now on the only non-zero terms will be NN and NNN hopping $t_{i,i+1}^\sigma \equiv t $ and $t_{i,i+2}=t_2$, as well as non-zero NN and NNN ring exchange $K_{i,i+1} \equiv K$ and $K_{i,i+2}=K_2$. For concreteness, we concentrate on the choice $t_2/t = K_2/K = 0.2$, corresponding to the leading terms of a power-law behavior with $\nu \approx 2.3$. We set $t=1$.

Finally, we also need to pay attention to the sign of each term, which can be selected through the choice of phonon branch: Coupling to transverse phonons allows for implementing antiferromagnetic interactions \cite{Porras2004,Zhu2006}, and therefore, this branch is used to generate the ring-exchange terms (hence $K_{ij}>0$). Coupling to longitudinal phonons allows for ferromagnetic interactions and, therefore, is used to generate the hopping terms (hence $-t_{ij}<0$). The selection of the phonon branch is done through the wave vector difference $\Delta k$ between two Raman lasers, and the desired choice is obtained through the laser arrangement depicted in Fig.~\ref{fig:scheme}(c).
In this arrangement, both couplings, $\{\alpha,\beta\}=\{-,0\}$ and $\{\alpha,\beta\}=\{+,0\}$, use the longitudinal phonon branch. To avoid interference effects between both couplings, different detunings will be needed, $\delta_{-0}^{\mathsf{m}} \neq  \delta_{0+}^{\mathsf{m}}$. The effective hopping amplitudes  $t_{ij}^\sigma=-J_{\sigma0}^{ij}$ in both legs can be made approximately equal if Rabi frequencies are chosen accordingly, $(\Omega_{0-}/\Omega_{0+})^2 = \delta_{0-}^{\mathsf{lcom}}/ \delta_{0+}^{\mathsf{lcom}}$, cf. Eq.~(\ref{Jij}). Here, the index ${\mathsf{lcom}}$ denotes the longitudinal center-of-mass mode, which is the dominant contribution to the $0+$ and $0-$ coupling in Eq.~(\ref{Jij}). The ring-exchange amplitude $K_{ij}=J_{-+}^{ij}$, generated by the coupling to transverse phonons, can be tuned fully independently.

\begin{figure}[t!]
    \centering
    \includegraphics[width=0.475\textwidth]{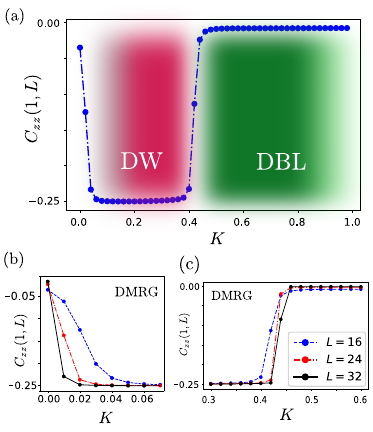}
    \caption{(a) The correlation $C_{zz}(1, L)$ is plotted as a function of $K$ using ED ($L=16$) at ${\cal P}=0$. The system passes from a state with a domain wall (DW) to a state without any local order (DBL). This transition persists with varying $L$, as observed from DMRG, as shown in (b) and (c). We demarcate separate phases on the plot to ease the visual representation of different phases.} 
    \label{fig:Trans}
\end{figure}

\section{Results} 
\begin{figure}[t!]
    \centering
    \includegraphics[width=0.45\textwidth]{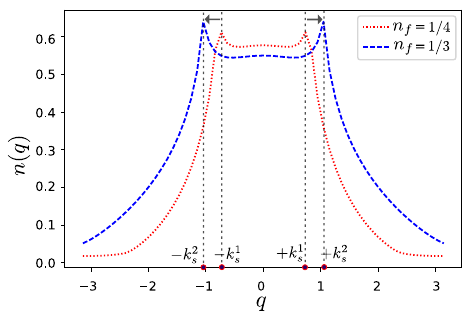}
    \caption{$n(q)$ is plotted for $K=0.6$, using DMRG for $L=60$. The absence of zero momenta peak and the presence of peaks at $k_s=\pi n_f$ are also consistent with the known features of such a phase. We see that changing $n_f $ from $1/4$ to $1/3$ shifts the peaks.
    }  \label{fig:DBL}
\end{figure}

\begin{figure}[t!]
    \centering
    \includegraphics[width=0.45\textwidth]{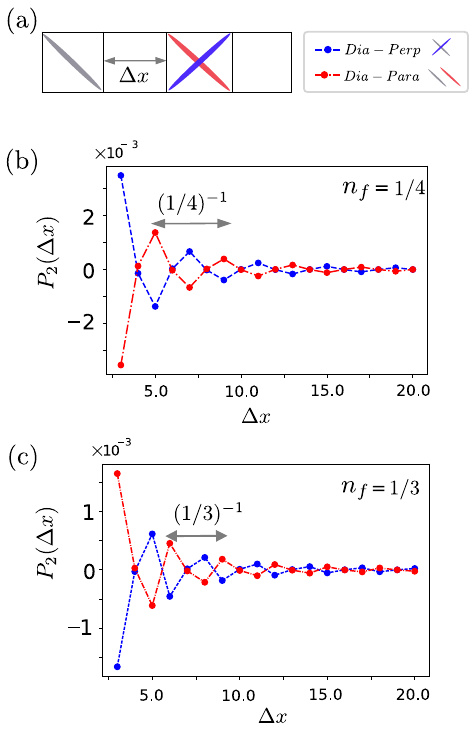}
    \caption{The $P_2 (\Delta x)$ for parallel diagonals: $Dia-Para$ ($\gamma =1, \eta=2 $ ) and perpendicular diagonals: $Dia-Perp$ ($\gamma =2, \eta=1 $ ) are opposite in sign and oscillate with a period $1/n_f$. In (a), we show the convention used for calculating the pair correlation, and in (b) and (c), results for different fillings are depicted. We observe changing $n_f$ changes the modulation in $P_2$ which is consistent with the shifting of peaks in FIG.~\eqref{fig:DBL}.}  \label{fig:P2}
\end{figure}

To study the model numerically via ED and DMRG\footnote{The DMRG sweeps are performed with the maximum allowed bond dimension varying between 1000-5000, depending upon the requirement, and truncation error is kept at $10^{-10}$.}, we concentrate on the filling $n_f=(\mathcal{N}_{+}+\mathcal{N}_{-})/(2L) = \sum_i \langle (S_z^i)^2 \rangle/(2L)$ to $1/4$, as well as the polarization  $\mathcal{P} = \mathcal{N}_{+}-\mathcal{N}_{-} = \sum_i \langle S_z^i \rangle$. We emphasize that our study is not specific to one particular filling, and although we begin with $n_f=1/4$, in what follows, we will report results for other fillings as well (as will be done in Fig. \ref{fig:DBL} and \ref{fig:P2}). In these definitions, $\mathcal{N}_{\pm}$ refers to the number of particles in the $|\pm \rangle$ legs, mapped onto the internal state of the ions via the spin-1 notation, $S_z^i=\tau^i_{++}-\tau^i_{--}$. 

The lowest energy in each polarization sector is plotted in Fig.~\ref{fig:Ens}(a) as a function of $K$. For $K<K_c \approx 0.44$, the true ground state is fully polarized ($|\mathcal{P}|=2Ln_f$). In this configuration, one leg remains empty, and hence, no ring exchange can occur, and the energy does not depend on $K$. Notably, in this regime, the lowest energy states in the other polarization sectors are all (at least approximately) degenerate. At $K=K_c$, a true level-crossing occurs, and an unpolarized state becomes the unique true ground state for $K>K_c$. 
Interestingly, this level crossing coincides with an avoided level crossing within the unpolarized sector, see Fig.~\ref{fig:Ens}(b). Therefore, even if the polarization is fixed to $\mathcal{P}$=0, as will be done below, an abrupt change of the system behavior can be observed at $K=K_c$. 

One quantity that provides evidence of the different physical properties in these two regions is the correlation function 
\begin{align}
C_{zz}(i,j) = \langle S_z^i S_z^j \rangle. \label{eq:Czz}
\end{align}
\begin{figure}
    \centering
    \includegraphics[width=0.475\textwidth]{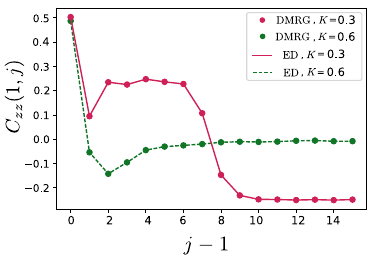}
    \caption{The correlator $C_{zz}(1,j)$ behaves differently in these two phases (i.e. $K<K_c$ and $K>K_c$). In the DW phase (red), positive correlations for $j \leq 8$ indicate that these rungs are polarized on the same leg as the first rung, whereas negative values for $j>8$ indicate opposite polarization, establishing a domain wall at $j=8$.
    } 
    \label{fig:TransSc}
\end{figure}
In Fig.~\ref{fig:Trans}, we use the long-distance behavior of the above-mentioned correlator, from one edge $i=1$ to the other edge $j=L$, to study the transition between these two regions as a function of $K$. 
%The specifics of these regions will become more discernible as we proceed further. 
We concentrate on the fully unpolarized sector $\mathcal{P}$=0, as the true ground state for the Bose metal phase resides here. Notably, the correlator $C_{zz}(1,L)$ vanishes for $K>K_c$, but takes non-zero negative values in an intermediate regime. 
This hints at the formation of a domain wall in the intermediate regime, with particles on opposite sides of the system being polarized on opposite legs. The domain wall picture is further evidenced by 
Fig.~\ref{fig:TransSc}, where we plot $C_{zz}$ as a function of distance $j-1$ from one edge, considering the lowest energy state in $\mathcal{P}$=0 at $K=0.3<K_c$ and $K=0.6>K_c$. For $K=0.3$, this quantity features the domain wall, with the left half of the system being polarized in one leg and the right half being polarized in the other leg. It shows the distribution of hardcore bosons throughout the chain if the first site, on the left, is occupied. In terms of the two-leg ladder system, the changing polarization pronounced by the sign change in $C_{zz}$ of Fig.~\ref{fig:TransSc}, refers to correlations developed between sites on different legs.
 The domain-wall picture also provides an intuitive explanation for the double degeneracy of the DW state, seen in Fig.~\ref{fig:Ens} (b). We may associate the energy gap to the fully polarized state with the energy cost of such a domain wall. 
 
We note that the correlations associated with the domain wall order are classical, e.g., they are of the form $\langle S_z^i \rangle \langle S_z^j \rangle$. In the appendix Sec.~\ref{Sec:SzConn}, we show that also after removing classical correlations by switching to the connected correlation function $C^{conn}_{zz} (\mid i-j \mid ) = C_{zz}(i,j) - \langle S_z^i \rangle \langle S_z^j \rangle$ can differentiate between the two regimes, through different scaling behavior with distance $|i-j|$. However, as $C^{conn}_{zz}$ vanishes at large distances for both phases, this quantity does not explicitly capture the domain wall configuration. 

 For very small values of $K$, the domain-wall picture does not apply in finite systems, as seen in Fig.~\ref{fig:Trans}(a). In this regime, the quasi-degeneracy of many states leads to different behavior. However, the data in Fig.~\ref{fig:Trans}(b) suggest that the critical $K$ of this regime tends to zero when the system size is increased. In contrast, $K_c$ at the boundary to the DBL phase turns out to be independent of the system size, see Fig.~\ref{fig:Trans} (c).

Importantly, the ground state for $K>K_c$ carries clear signatures of a $d$-wave Bose liquid (DBL). Two quantities of utmost interest in defining the DBL phase are: ({\it i}) the momentum distribution function
\begin{align}
n(q) = \sum_{j_1,j_2} \sum_{\sigma=\uparrow,\downarrow} \exp (-i q (j_1-j_2)) \langle a^{\dagger}_{j_1,\sigma} a_{j_2,\sigma} \rangle /L,
\label{eq:nq}
\end{align}
which is plotted in Fig.~\ref{fig:DBL} for a relatively large system at $K=0.6>K_c$, and ({\it ii}) the pair correlation between diagonal sites of the two-leg ladder, given by 
\begin{align}
P_2(\Delta x) = \langle a^{\dagger}_{1,\uparrow} a^{\dagger}_{2,\downarrow} a_{\Delta x + \gamma,\uparrow} a_{\Delta x +\eta,\downarrow}  \rangle ,
\label{eq:P2}
\end{align}
which in illustrated Fig.~\ref{fig:P2} for different filling factors. The choice ($\gamma =1, \eta=2 $) denotes correlation between two parallel diagonals and ($\gamma =2, \eta=1 $) implies correlation between two perpendicular diagonals. The convention is schematically described in Fig.~\ref{fig:P2}(a). The oscillatory correlation pattern captures the $d$-wave nature of the DBL.

Notably, there is no zero momentum peak in Fig.~\ref{fig:DBL}, implying the absence of boson condensation. The new peaks at finite momenta denote the emergence of new singularities in the momentum distribution, revealing metallic features of the many-body ground state. We further note that the position of the $n(q)$ peaks as well as the oscillation period in $P_2 (\Delta x)$ are both connected to the filling of the ladder, as has already been worked out in Ref.~\cite{Sheng_2008}. Hence, we observe the development of a new scale in the system getting reflected both in single-particle ($n(q)$) and two-particle ($P_2$) correlators.

Apart from the quantities evaluated and discussed above, there are additional observables that can be used to characterize the DBL and the DW-DBL transition: This includes the local magnetization $\langle S_z^j \rangle$, the connected correlation function $C_{zz}^{conn}$, or the structure factor. The latter becomes useful for determining the liquid-like features of the DBL phase. We have dedicated an entire appendix to a detailed study of all these quantities. We refer the interested reader to Sec.~\ref{Sec:SzConn}.

\section{Discussion and Outlook}
In this paper, we have proposed to use a chain of three-level ions to implement a model of bosons on a two-leg ladder with hard-core conditions on each rung. The setup allows us to explore the effect of tunable ring exchange interactions $K$, and we have shown that they produce a transition from a polarizing regime at small $K$ to a regime with features of a $d$-wave Bose liquid at larger $K$.

The recent advancements in trapped ion experiments \cite{Monroe2021} present themselves as an ideal platform for realizing the physics under study. However, we note that trapped ion systems feature all-to-all couplings, whereas our study has focused on a model with only nearest- and next-to-nearest neighbor couplings to facilitate DMRG calculations. Using ED, we have explicitly checked that the reported behavior qualitatively persists also in the presence of all-to-all interactions, as long as they decay with a power-law exponent of two or larger. We direct the readers to Sec.~\ref{Sec:LR} for a detailed study on DBL transition with all-to-all hopping.
Such a decay can easily be achieved when the coupling is transmitted by transverse phonons, as used here to implement the ring exchange term, however, it might be hard to achieve with longitudinal phonons, due to a larger bandwidth of the spectrum \cite{Zhu2006}. Since we rely on longitudinal phonons to implement the hopping terms through antiferromagnetic couplings, additional tuning might be necessary to achieve sufficiently fast decay. One possibility here is Floquet engineering techniques; see for instance Ref.~\cite{Christ2023,Rajibul2024}.
A flexible alternative to the analog implementation of the model is the simulation using a digital qutrit quantum computer \cite{Ringbauer_2022}. 

Finally, let us also comment on the possibility of implementing the hopping term via ferromagnetic interactions transmitted through transverse phonons. The negative sign of the hopping amplitude amounts to a $\pi$-flux, which, quite interestingly, is found to stabilize the DBL regime even in the absence of ring exchange. Although such a setup is experimentally less demanding, in this paper, we have focused on the implementation with $t>0$, which features a DBL transition induced via ring exchange interactions.

\acknowledgments{ 
The authors thank Martin Ringbauer and Claire Edmunds for reading our manuscript and for providing helpful feedback on experimental aspects and implementation.
We acknowledge the financial support received from the IKUR Strategy under the collaboration agreement between the Ikerbasque Foundation and DIPC on behalf of the Department of Education of the Basque Government. T.G. acknowledges funding by the Department of Education of the Basque Government through the project PIBA\_2023\_1\_0021 (TENINT) as well as the Joint Research Agreement between BasQ and IBM, by the Agencia Estatal de Investigación (AEI) through Proyectos de Generación de Conocimiento PID2022-142308NA-I00 (EXQUSMI). This work has been produced with the support of a 2023 Leonardo Grant for Researchers in Physics, BBVA Foundation. The BBVA Foundation is not responsible for the opinions, comments, and contents included in the project and/or the results derived therefrom, which are the total and absolute responsibility of the authors.
E.R. acknowledges support from the BasQ strategy of the Department of Science, Universities, and Innovation of the Basque Government. E.R. is supported by the grant PID2021-126273NB-I00 funded by MCIN/AEI/ 10.13039/501100011033 and by ``ERDF A way of making Europe" and the Basque Government through Grant No. IT1470-22. This work was supported by the EU via QuantERA project T-NiSQ grant PCI2022-132984 funded by MCIN/AEI/10.13039/501100011033 and by the European Union ``NextGenerationEU''/PRTR. This work has been financially supported by the Ministry of Economic Affairs and Digital Transformation of the Spanish Government through the QUANTUM ENIA project called – Quantum Spain project, and by the European Union through the Recovery, Transformation, and Resilience Plan – NextGenerationEU within the framework of the Digital Spain 2026 Agenda.}

\bibliographystyle{quantum}
\bibliography{bib}

\onecolumn 
\appendix
\section{Supplemental data on the DBL transition} \label{Sec:SzConn}

\begin{figure} [h!]
    \centering
    \includegraphics[width=0.9\textwidth]{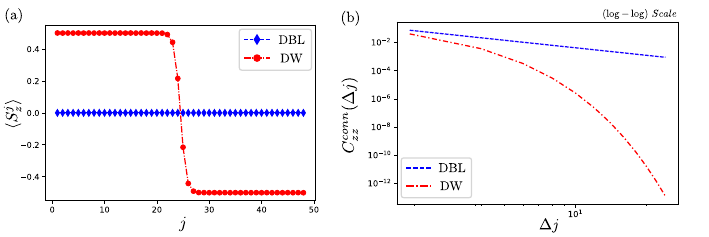}
    \caption{Both the panels are plotted with $L=48$, where for DBL we use $K=0.6$ and for the DW we use $K=0.3$. In (a) we show the local magnetization. It is evident that for DBL there is no local order. However, for DW there is a finite local order. In (b) we plot the decay of the connected correlator in the bulk, using a $\log-\log$ scale. The $C_{zz}^{conn}$ decays faster following an exponential behavior for the DW, whereas it is a power law for DBL.}  \label{fig:SzConn}
\end{figure}

This section explicitly shows three different correlations to distinguish DBL from DW. The results below complement the principal findings reported in the main text. In Fig.~\ref{fig:SzConn} we show the local magnetization. 

\begin{figure}[h!] 
    \centering
    \includegraphics[width=0.475\textwidth]{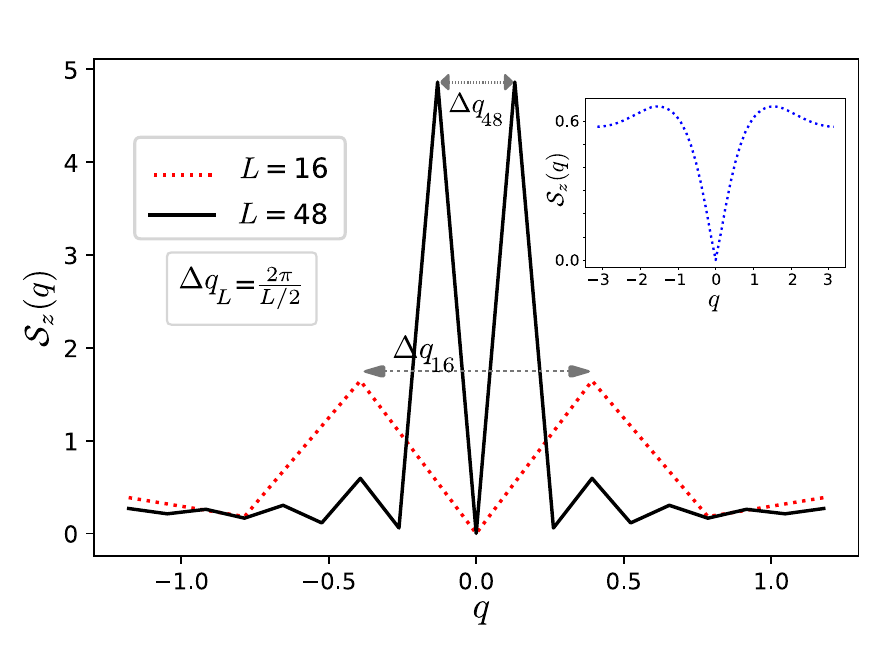}
    \caption{$\mathcal{S}_z(q)$ computed using DMRG.  At $K=0.3$, a comparison between $L=16,48$ shows the persistence of sharp peaks with $\Delta q_L$ wave vector. In the inset, we see data for $L=48$ at $K=0.6$. The smoothness of $\mathcal{S}_z$ suggests an absence of any true long-range order. This is a characteristic of the DBL phase.  }  \label{fig:SFq}
\end{figure}

As an indicator of the liquid-like behavior of the system for $K>K_c$, we have also studied the $z$-component of structure factor defined as, 
\begin{align}
\mathcal{S}_z(q) = \sum_{j_1,j_2} \exp (-i q (j_1-j_2)) \langle S_z^{j_1} S_z^{j_2} \rangle /L ,
\label{eq:Sfq}
\end{align} 
which has sharp peaks for $K<K_c$, but gets smoothened for $K>K_c$. Corresponding data is shown in Fig.-\ref{fig:SFq}. In Fig.~\ref{fig:Trans}, we observed that for $K<K_c$, the correlation $C_{zz}(1,j)$ changes sign at $L/2$, signifying the existence of two regions with different polarizations. The length of such domains is $L/2$ (which is consistent with findings from local magnetization as shown later in Fig.~\ref{fig:SzConn}(a)). This observation is further substantiated by the appearance of sharp peaks at wave vector $q=2\pi/(L/2)$, as shown in Fig.~\ref{fig:SFq}. In the inset of the same figure, one can find out how no such peak exists in the DBL phase at $K>K_c$.

The DBL does not show any local order, and the magnetization $\langle S_z^{j} \rangle$ vanishes (see Fig.~\ref{fig:SzConn}(a)). However, in the DW phase, it is possible to construct states with non-zero local magnetization. In this context, one has to note that the unpolarized system is doubly degenerate in the DW phase, and the two states have opposite magnetization. However, it is possible to choose a linear combination of both states that features the domain wall. This is shown in the red curve of Fig.~\ref{fig:SzConn}(a), where $\langle S_z^{j} \rangle$ flips the sign in the center of the system. This is equivalent to particles being polarized on two opposite segments on each of the legs, as mentioned in the main text. In DBL, there is no local order, and this quantity should be zero throughout the system. 

We further analyze the connected correlators in the bulk of the system, defined as
\begin{align}
C_{zz}^{conn} (\Delta j) =\langle S_z^{L/2+\Delta j} S_z^{L/2-\Delta j} \rangle   - \langle S_z^{L/2+\Delta j} \rangle \langle S_z^{L/2-\Delta j} \rangle.
\label{eq:Czzconn}
\end{align}
For a classical domain wall state with non-zero magnetization, the connected correlators go to zero at large distances, as they also do in the DBL phase. Nevertheless, the connected correlators can differentiate clearly between a DW and the DBL state through their scaling behavior, shown in Fig.~\ref{fig:SzConn}(b): As expected, the decay turns out to be exponentially fast in the domain wall (DW) state, compared to an algebraic decay in the DBL 
%As $C_{zz}^{conn}$ vanishes at large distances, the quantity $C_{zz}(i,j)$ becomes a reliable marker to differentiate between the DW and the DBL. In particular, $C_{zz}(1,L)$ unambiguously distinguishes DBL from DW, as it is strictly zero in only DBL and finite in the other. We extend this observation to exploit the difference between $C_{zz}$ and $C_{zz}^{conn}$ in favor of numerical analysis since $\langle S_z^{j}\rangle$ is zero only for DBL. 
The power law scaling of the connected correlator is suggestive of strong quantum fluctuations present in the DBL.
%and not sufficient to characterize the phase.

\section{Study of full long-range model} \label{Sec:LR}
The physical implementation of the proposed set-up in the trap gives rise to long-range terms, both for the exchange interactions and the hopping. The strength of these single and two-body processes can be approximated as a power law decay, $t_{ij} \sim |i-j|^{-\alpha}$ and $K_{ij} \sim |i-j|^{-\alpha}$. The exponent $\alpha$, which depends on the detuning of the Raman coupling, controls the extent of long-range processes over the scale of the lattice constant. 

The DMRG technique is not reliable for the analysis of long-range interaction. Therefore, we have resorted to ED, which is limited to relatively small systems, $L \leq 16$. A figure of merit that is capable of tracking the DBL transition in small systems is $\Delta n \equiv n(0) - \max(n(q))$, the difference between the momentum distribution at $q=0$ and peak value. As discussed in the main text, the onset of the DBL phase manifests itself through a splitting of the zero momenta peak in $n(q)$, giving rise to non-zero values of $\Delta n$.

\begin{figure} [h!]
    \centering
    \includegraphics[width=0.95\textwidth]{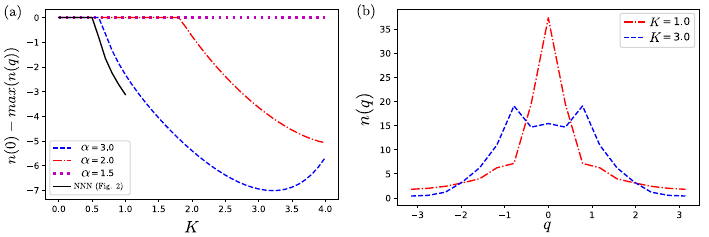}
    \caption{(a) The quantity $\Delta n =n(0) - {\rm max}(n(q))$ is plotted vs. $K$ to identify DBL formation through the characteristic features of $n(q)$, using ED for a chain with $L=16$. The zero value of the quantity implies that the maximum of $n(q)$ corresponds to $q=0$. A finite negative value, with increasing $K$, denotes the appearance of a momenta peak at a non-zero value, indicating peak splitting which is typical of the DBL phase. We plot the same function for different $\alpha$. For
    $\alpha \geq 2$, the DBL transition is observed, whereas for $\alpha =1.5$ it is not. We have also shown the $\Delta n$, corresponding to the data set of Fig. 2 of the main text, for the sake of completeness. (b) We show $n(q)$ for $\alpha=2$, to illustrate the splitting of peaks at sufficiently large ring exchange ($K=3.0$, red dashed-dotted line), in contrast to a single peak for sufficiently small ring exchange ($K=1.0$, blue dashed line).}  \label{fig:ED_Sp}
\end{figure}

In Fig.~\ref{fig:ED_Sp}, $\Delta n$ is plotted vs. $K$ for three different exponents $\alpha$. 
For $\alpha \geq 2$, clear signatures of the DBL are visible, and we conclude that a long-range system of coupling parameters with quadratic or faster decay will exhibit a DBL transition, shifted towards larger $K$ as $\alpha$ is decreased. However, in the data for $\alpha = 1.5$, no signature of DBL is obtained. 

We should also mention that due to restrictions on the Hilbert space dimension, we are unable to go beyond $L=16$. Hence, the shown results are not devoid of possible finite-size effects. 
In this context, we note that the critical value obtained in Fig.~\ref{fig:ED_Sp}(a) from the splitting of the momentum distribution peak is slightly shifted (by $~0.1$ on the scale of $K$-axis) for other markers of the DBL phase, such as crossings of energy levels from different polarization sectors, or correlation data $C_{zz}$. We associate this tiny mismatch with the small size of the system, supporting the need for DMRG analysis for a more precise determination of the phase boundary.

\section{Raman coupling and effective Hamiltonian} 
\label{Sec:EffH}
For a formal description of the Raman coupling, we go into the rotating frame of $H_0$ and apply a rotating wave approximation. With this, and under the Lamb-Dicke assumption, i.e. $e^{i \Delta k x}\approx 1 + i \Delta k x$, the coupling between levels $\alpha,\beta \in \{-,0,+\}$ can be expressed as
\begin{align}
\label{h}
    h_{\alpha \beta}(t) = \frac{i \Omega_{\alpha\beta}}{2} \sum_{i,\mathsf m}
 \eta_{{\mathsf m},i}  
e^{i(\mu_{\alpha\beta} -\omega_{\mathsf m})t} 
b_{\mathsf m}^\dagger \tau^i_{\alpha\beta} + {\rm h.c.},
\end{align}
with  $\Omega_{\alpha\beta}$ being the Rabi frequency, $\eta_{{\mathsf m},i}$ the Lamb-Dicke parameter for mode $\mathsf m$ at ion $i$, and $b_{\mathsf m}^\dagger (b_{\mathsf m})$ the creation (annihilation) operator of this mode. 
Each coupling has an individual beat-note $\mu_{\alpha\beta} = \omega_{{\rm com},\alpha\beta} + \delta_{\alpha\beta}$, detuned by $\delta_{\alpha\beta}$ from the center-of-mass (com) mode of the selected phonon branch, with frequency $\omega_{{\rm com},\alpha\beta}$. In total, we require three such coupling terms, $H(t)= h_{0-}(t) + h_{0+}(t) + h_{-+}(t)$, where $h_{0-}$ and $h_{0+}$ implement the hopping (green couplings in Fig.~\ref{fig:scheme}), and $h_{-+}$ produces the ring-exchange (yellow coupling in Fig.~\ref{fig:scheme}).

The effective Hamiltonian in Eq.~(\ref{Heff}) is obtained from a second-order Magnus expansion of the time-evolution operator $U_{\alpha\beta}(t,0)={\cal T} \exp[-\frac{i}{\hbar}
\int_0^t d\tau \ h_{\alpha\beta}(\tau)]$. Specifically, the second order of the Magnus expansion will give rise to non-oscillatory terms which are of the form $e^{-i H_{\rm eff} t/\hbar}$. Hence, these terms provide the notion of an effective, time-independent Hamiltonian $H_{\rm eff}$.

Up to the second order, the Magnus expansion of the evolution operator reads
\begin{align}
U_{\alpha\beta}(t,0) \approx \exp\Bigg\{ -\frac{i}{\hbar} \Bigg[ \int_0^t d\tau h_{\alpha\beta}(\tau) -\frac{1}{2}\int_0^td \tau_2\int_0^{\tau_2} d\tau_1  
[h_{\alpha\beta}(\tau_2),h_{\alpha\beta}(\tau_1)]\Bigg]\Bigg\}.
\end{align}
In this expansion, we are only interested in terms that are linear in $t$, as they will establish the notion of an effective time-independent Hamiltonian. To find these terms, we first note that each term in $h_{\alpha\beta}(\tau)$ is of the form $b_{\mathsf{m}}^\dagger \tau_{\alpha\beta}^i e^{i\delta_{\alpha\beta}^{\mathsf{m}}t}$ or $b_{\mathsf{m}} \tau_{\beta\alpha}^i e^{-i\delta_{\alpha\beta}^{\mathsf{m}}t}$. From this observation, we immediately see that all first-order terms in the Magnus expansion will maintain oscillatory behavior. However, in second-order, there are integrals of the form $\int_0^td \tau_2\int_0^{\tau_2} d\tau_1 e^{i(\delta_2 \tau_2 - \delta_1 \tau_1)}$, which for $\delta_1=\delta_2\equiv \delta$ give rise to a $t$-linear term, as well as an oscillatory term and a constant term:  
\[
\int_0^td \tau_2\int_0^{\tau_2} d\tau_1 e^{i \delta (\tau_2 - \tau_1)}= it\frac{1}{\delta} + \frac{1-e^{i\delta t}}{\delta^2}.
\]
Thus, to collect the $t$-linear terms, we have to account only for pairs of phonon creation/annihilation processes at equal detunings. This observation justifies that the interference of different Raman couplings with different detunings does not affect the effective Hamiltonian. 
Hence, we indeed can treat each coupling term $h_{\alpha\beta}(t)$ independently from each other. We explicitly stress the fact that the time scale $t$ of interest is
defined by the inverse of the effective coupling strength $J_{\alpha\beta}^{ij}$, see Eq.~(\ref{Jij}). This turns out to be much larger than $\delta^{-1}$, such that in this regime of $t$, ${\rm Im}(it\frac{1}{\delta} + \frac{1-e^{i\delta t}}{\delta^2})$ behaves essentially linear in $t$, in contrast to the quadratic behavior for $t \to 0$.

Finally, by evaluating the commutator, we find that the second-order integral is given by
\begin{align}
\int_0^t d\tau_2 \int_0^{\tau_2} d\tau_1   [h_{\alpha\beta}(\tau_2),h_{\alpha\beta}(\tau_1)] &=
it \sum_{i,j,{\mathsf{m}}} \frac{\Omega_{\alpha\beta}^2}{4\delta_{\alpha\beta}^{\mathsf{m}}} \times
\eta_{\mathsf{m},i} \eta_{\mathsf{m},j} [2n_{\mathsf{m}}(\tau_{\alpha\alpha}^i-\tau_{\beta\beta}^i) \delta_{ij} - \tau_{\alpha\beta}^i\tau_{\beta\alpha}^j - \tau_{\alpha\beta}^j\tau_{\beta\alpha}^i] \nonumber \\ &~~~ + {\rm oscillating \ and \ constant \ terms}, 
\end{align}
where $n_{\mathsf{m}}=b^\dagger_{\mathsf{m}}b_{\mathsf{m}}$.
The $t$-linear term leads to the effective Hamiltonian given in Eq.~(\ref{Heff}), and the coupling $J_{\alpha\beta}^{ij}$ given in Eq.~(\ref{Jij}) and the potential $V_{\alpha\beta}^i$ given in Eq.~(\ref{Vi}) can be read off from this expression.

\end{document}